\documentclass[conference]{IEEEtran}

\usepackage{color}
\usepackage{epsf}
\usepackage{times}
\usepackage{epsfig}
\usepackage{epstopdf}
\usepackage{graphicx}
\usepackage{algorithm}
\usepackage{algorithmic}
\usepackage{amsmath}
\usepackage{amssymb}
\usepackage{amsxtra}
\usepackage{multirow}
\usepackage{mathtools}
\usepackage{subcaption}
\usepackage{multirow}
\usepackage{comment}
\usepackage[normalem]{ulem}
\usepackage{tabu}
\usepackage[dvipsnames]{xcolor}
\usepackage{multicol}
\usepackage{cite}
\usepackage{caption}
\begin{document}

\title{A Photo-Based Mobile Crowdsourcing Framework for Event Reporting}

\author{\IEEEauthorblockN{Aymen Hamrouni$^{1,2}$, Hakim Ghazzai$^2$, Mounir Frikha$^1$, and Yehia Massoud$^2$}
\IEEEauthorblockA{\small $^1$ Higher School of Communications of Tunis, University of Carthage, Tunisia\\
Email: \{aymen.hamrouni, m.frikha\}@supcom.tn\\
$^2$School of Systems  \& Enterprises, Stevens Institute of Technology, Hoboken, NJ, USA \\
Email: \{ahamroun, hghazzai, ymassoud\}@stevens.edu}
}

\maketitle
\thispagestyle{empty}

\begin{abstract}
\boldmath{Mobile Crowdsourcing (MCS) photo-based is an arising field of interest and a trending topic in  the domain of ubiquitous computing. It has recently drawn substantial attention of the smart cities and urban computing communities. In fact, the built-in cameras of mobile devices are becoming the most common way for visual logging techniques in our daily lives. MCS photo-based frameworks collect photos in a distributed way in which a large number of contributors upload photos whenever and wherever it is suitable. This inevitably leads to evolving picture streams which possibly contain misleading and redundant information that affects the task result. In order to overcome these issues, we develop, in this paper, a solution for selecting highly relevant data from an evolving picture stream and ensuring correct submission. The proposed photo-based MCS framework for event reporting incorporates (i) a deep learning model to eliminate false submissions and ensure photos credibility and (ii) an A-Tree shape data structure model for clustering streaming pictures to reduce information redundancy and provide maximum event coverage. Simulation results indicate that the implemented framework can effectively reduce false submissions and select a subset with high utility coverage with low redundancy ratio from the streaming data.

}
\end{abstract}\vspace{0.1cm}


\section{Introduction}
\label{Sec1a}

Today, nearly 57\% of the world population lives in urban areas, a percentage that is expected to increase to 68\% by 2050~\footnote{A.  Lardier, ``Report:  Two-Thirds  of  Worlds  Population  Will  Live  in Cities by 2050,'' https ://goo.gl/TpiRNG/, 2018}. This means that urbanization combined with the overall population growth could add another 2.5 billion people to urban areas by that time. This will certainly lead to many social concerns regarding pollution, traffic congestion, and public safety. The existing solutions to deal with these matters, mainly Internet-of-things (IoT) solutions, share the idea of setting up as many fixed sensors as possible across cities and connect these sensors to citywide networks for uploading sensing data to cloud servers~\cite{8663367,8053134}. However, the deployment, management, and maintenance of fixed sensors are expensive and time-consuming. In addition, the number of sensors to be deployed in different areas should be depending on the population density of each area. Since the latter is time-sensitive as people are always on the move, it is hard to determine the appropriate sensors' density in each area that satisfies the demands at all times.

Mobile Crowdsourcing (MCS) can be a potential candidate to complement IoT technology and enhance smart city applications. MCS utilizes the power of mobile devices to accomplish specific sensing and data collection tasks without requiring predeployed dedicated infrastructure. It refers to the technology that takes advantage of wearable's and mobile phones' features like built-in sensors to collect and process the data of people and surrounding environments. Unlike fixed sensors, these devices are infrastructure-less and consequently more flexible. Typically, MCS is composed of three parties: task requesters, task workers, and a platform made available in the cloud. When a task requester find difficulties in collecting certain information, he/she can initiate a crowdsourcing task describing his/her problems and then, announce it via the platform to the crowd. The platform will be in charge in selecting, according to certain criteria, the group of appropriate contributors which can deliver satisfying results.

In many crowdsourcing applications, the task requester may sometimes need to collect photos of specific ongoing events and keep track of any updates. It is trivial to assume that the event report requester wants an outcome that has high utility coverage with low redundancy ratio and only contains error-free submissions. This could be, for example, the case of a local authority tracking a public safety concern, a weather company investigating abnormal meteorology behavior, or a newspaper or  media agency covering a real-time event and so on. To satisfy these needs, a two-step process must be put in place: (i) analyze the submitted photos to discard false submissions and (ii) preserve only the ones that maximizes the event coverage (i.e. the most diverse data associated to the same event).

Lately, MCS has been a trending research topic in the field of ubiquitous computing. It has attracted substantial attention from the smart cities and urban computing research communities and it is rapidly gaining popularity. Chen et al.~\cite{7807311} proposed a constraint-driven data selection model for mobile photographing from the perspectives of both data sensing and transmission. They presented approaches to collect MCS data with low redundancy. Zhang et al.~\cite{6871668} exploited the 4W1H—a four-stage life cycle to characterize the MCS process. Typical MCS applications include monitoring the pollution of rivers~\cite{Kim:2011:CWP:1978942.1979251}, traffic reporting in urban areas~\cite{Koukoumidis:2011:SLM:1999995.2000008}, reposting and sharing fliers distributed in residential communities~\cite{Guo:2014:FCP:2638728.2638730}, and monitoring prices of goods in the market~\cite{mobishop}.

Most of these MCS applications mentioned earlier are designed for only one single task where one type of data, for example, photos of potholes, is collected. They usually use specific photo selection criteria according to the application goal to discard redundant and low-quality photos (e.g., too dark or motion-blurry). Different from these existing MCS applications, a multitask MCS framework approach should be defined to check each input for trustworthiness (i.e. false submissions) and, at the same time, maintain only the data subset that has the maximum event coverage using multiple constraints defined by the task requesters.

In this paper, we propose a generic MCS framework for event reporting using photos as collected data. Selected workers, that can be human being, e.g., people using their smartphones, smartwatches, or cameras, or machines, e.g., CCTV, drones,  are asked to collect and submit photos about a specific ongoing event happening in a given geographical location and at a certain time. The proposed framework handles both the photo selection problem for a high event coverage using an A-tree shape hierarchical data structure and eliminates false submission by using a deep learning model that eliminates wrong and inaccurate reports.
Simulation results using our proposed framework confirms our proposed model and shows that our solution can reduce incorrect submissions and remove data redundancy.

\section{Event Reporting MCS Framework}\label{sec3}

\begin{figure}[t]
        \includegraphics[width=0.5\textwidth]{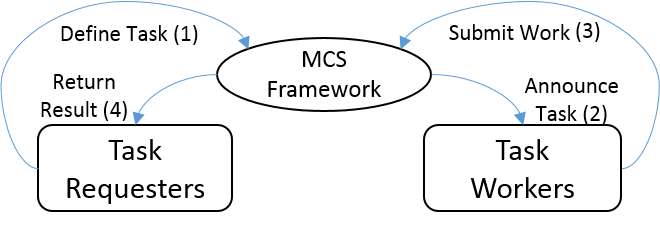}
    \caption{Typical MCS Workflow.}
    \label{d}\vspace{-0.4cm}
\end{figure}

A typical MCS framework interacts with two general actors: task requesters and task workers, as highlighted in Fig.~\ref{d}. It can be divided into four-stage process:
1) a task initiation stage to define and announce tasks to future-selected contributors,
2) a task execution stage during which the worker is supposed to collect and submit requested data,
3) a data aggregation phase that processes and filters submitted data using task requester constraints, and 
4) a result handover stage for delivering the post-processed data to the task requester.

\begin{figure}[t]
%\vspace{-0.5cm}
        \includegraphics[scale=1.1]{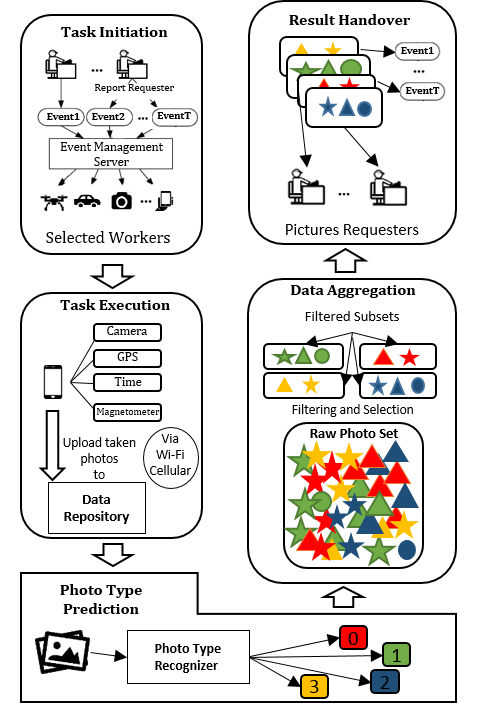}
        \caption{MCS event reporting framework high-level architecture.} 
    \label{a}\vspace{-0.4cm}
\end{figure}

In this paper, we ameliorate this architecture by enhancing the Data Aggregation (DA) stage and adding a Photo Type Prediction (PTP) phase as illustrated in Fig.~\ref{a}. In other words, we focus our study on the stages (3) and (4) mentioned in Fig.~\ref{d}. The remaining steps of the global framework will be investigated in the future extension of this work. The proposed PTP phase is responsible for mainly two important tasks: 1) predicting the type of each worker input photo and 2) eliminating false submissions meaning that the photos that do not correspond to the event in question will be discarded. This second task can be also seen as defining the event type if it has not been already predefined by the task requester. In this case the event type will be determined based on the majority of worker's submissions. This is the case of an offline scenario where we assume that the majority vote is credible and the PTP phase will occur only once the task deadline is over (this case is fully explained in the following section).

At the task initiation stage, data requesters outline their tasks with different requirements to the task management server. The latter which is charged of assigning them to appropriate workers will select and announce the task and its constraint to the most suitable contributors while considering the requirements pre-defined by requesters (e.g., the closest ones to the event, the lowest reward seeking, etc). At the task execution stage, the chosen workers take photos according to the task requirements and upload them to the crowdsourcing server. As the server receives photos uploaded by distributed workers intermittently, the photo stream inevitably contains redundant and possibly false information. As such, after going through the PTP stage, during which false submissions are discarded, the DA phase will filter the uploaded photos in order to remove redundant informations. Finally, in the result handover stage, the data repository with the appropriate photos and the event type is made available to the data requester upon task completion.

Since inaccurate and duplicate photos can lead to unwanted data traffic in MCS applications, a possible approach to solve this problem can be by uploading the photo's thumbnail and its related contextual informations to be analyzed beforehand at the server level using the MCS task constraints. To sum up, the picture stream goes through the following two steps: 1) running the captured photo of event through an algorithm to check their credibility and eliminate false submission (PTP phase) and 2) once verified, the photo is uploaded to the server side and goes through the DA phase for duplication checking. The ones that fails to pass the verification process will be automatically discarded.

\subsection{Photo Type Prediction Phase}
The purpose of the PTP phase is to (1) reduce the number of false submissions if the event type is pre-known by the task requester and (2) report the event type otherwise. This can be modeled as a photo recognition deep learning problem with $M$ classes where $M-1$ classes describe different event types that can be handled by the crowdsourcing platform (e.g. wildfire, incidents of mass violence, etc.) and a class for normal everyday photos (e.g. landscapes, selfies, etc). In this context, we propose two ways to execute this process: 1) an online scenario where, during the photo collection process, the photo type predictor ignores false submission assuming a prior knowledge about the nature of the reported event, i.e., the task requester informs the platform about the event type when initiating the task. The second way is 2) an offline scenario where, the platform needs to wait until the task deadline to filter redundant photos. Afterwards, the platform determines the nature of the events happening at the locations specified by the task requester. The filtering decision is based on the average of submitted photos. We want to point here that the online scenario requires that the event type is known beforehand by the task requester. However, the offline scenario may assume a zero knowledge about the event to be reported.

\subsection{Data Aggregation Phase}
It is important to select a subset of photos that maximizes the event coverage and contains small if not zero redundancy level. Existing MCS photo-based applications have different criteria for DA. Table~\ref{relatedapp} summarizes few of these MCS applications and constraints used in their data selection.

In order to select diverse photos for obtaining wider perspectives, MCS photo-based applications usually use one of the following two criteria to measure photos’ similarity. One is to assess photos’ visual contents,
e.g., FlierMeet \cite{Guo:2014:FCP:2638728.2638730}. The other is to assess photos’ semantic contents based on photographing contexts of photos, e.g., InstantSense \cite{7524481}.

The goal of the DA phase is to reduce submitted photos’ redundancy rate before result handover. Since the DA comes after the PTP step, it is only natural to assume that the photos going through the DA phase are verified and relative to the event. This means two possible scenarios: 1) In the online scenario, the data aggregator step will handle all photos that their types, according to the PTP phase, matches the task requester expectations. 2) In the offline scenario, the DA phase will handle all photos that their types were submitted the most.

There are different methods to compute photos’ similarity distance according to one or two features of the photo for redundancy detection. In this paper, our framework will assess photos’ similarity based on both, the combinations of photos’ visual and semantic features, and the photo proprieties (e.g. when and where the photo were taken).

\begin{figure}[t]
\begin{minipage}[h]{1\linewidth}
    \centering
    \vspace{0.2cm}
    \includegraphics[width=1\textwidth]{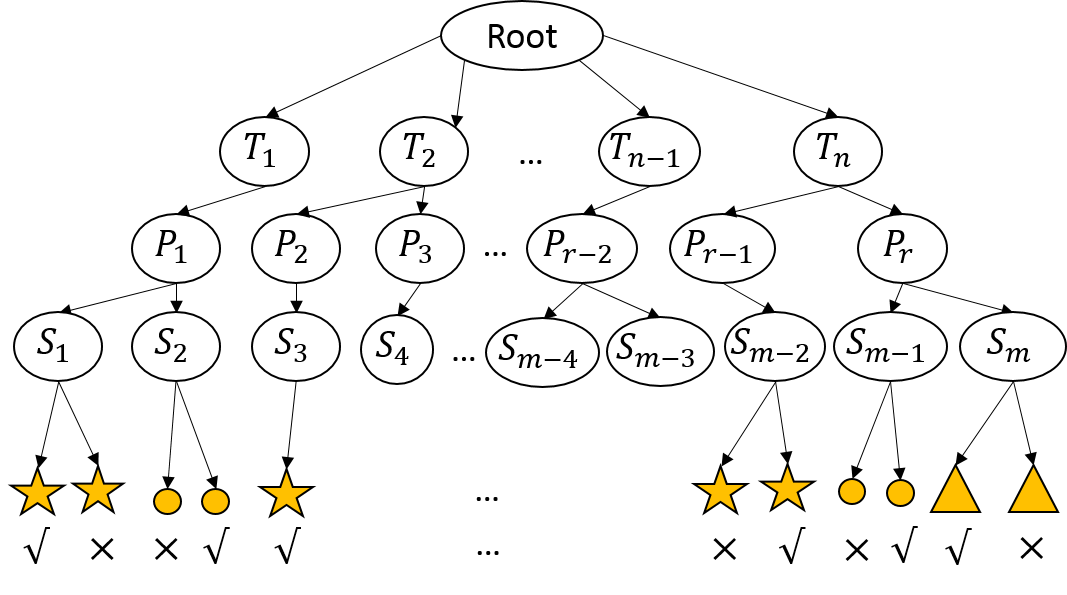}\vspace{-0.2cm}
    \caption{Example of an A-tree shape hierarchical structure of the data aggregator with $D=3$ layers: (T) for time, (P) for position, and (S) for photo similarity. }
    \label{f9}
\end{minipage}\vspace{-0.4cm}
\end{figure}

Our Data Aggregator is a $(D+2)$-layer hierarchical tree structure as shown in Fig.~\ref{f9}. Its root node is in the $0$-th layer and the classified photos are leafs within the $(D+1)$-layer. The $D$ layers in the middle represent the decision constraints that can be used to determine photos similarity (e.g. time, GPS location, etc). Each layer $d$ in $[1,D]$ represents one of those constraint. When the event report requester defines the task, he/she must specify the number and type of constraints to be used along with the metrics (e.g. time duration, distance, etc). Based on these parameters, the data aggregator decides whether photos are similar or not. We would like to point that in some cases the photo aggregator may keep redundant data in the tree. This can be, for example, the case when submitting similar photos taken at the same location but during different time instants and the time-difference is higher than the time duration threshold defined by the report requester. Also, we want to refer that the order of the $d$ layers do not impact the filtering process since the A-tree is sequence-independent.
\begin{table}[t]
\begin{center}
\caption{\label{relatedapp} Selection Criteria and Related MCS Applications}
\addtolength{\tabcolsep}{-4pt}
\resizebox{9cm}{!} {
\begin{tabular}{|c||c|}%
  \hline
 \textbf{Criteria } & \textbf{Applications} \\
  \hline
  \hline
\vtop{\hbox{\strut Multiple directions}\hbox{\strut Single direction }}
& 
\vtop{\hbox{\strut PhotoCity\cite{Tuite:2011:PTE:1978942.1979146}, InstantSense\cite{7524481}.}\hbox{\strut FlierMeet\cite{Guo:2014:FCP:2638728.2638730}, MobiShop\cite{mobishop}}
}
\\
  \hline
\vtop{\hbox{\strut Local}\hbox{\strut Global }}
& 
\vtop{\hbox{\strut WreckWatch\cite{White:2011:WAT:1997768.1997797}, InstantSense\cite{7524481}.}\hbox{\strut CreekWatch\cite{Kim:2011:CWP:1978942.1979251}.}} 
 \\
  \hline
\vtop{\hbox{\strut Small time slot}\hbox{\strut Large time slot }}
& 
\vtop{\hbox{\strut WreckWatch\cite{White:2011:WAT:1997768.1997797}, InstantSense\cite{7524481}.}\hbox{\strut PhotoCity\cite{Tuite:2011:PTE:1978942.1979146}.}}
\\
 \hline
 
\end{tabular}
}
\end{center}\vspace{-0.2cm}
\end{table}

Initially  empty,  the  A-Tree  will  grow by a  new leaf with each coming photo. The position of a new leaf in the A-Tree is resolved by matching the incoming photo's data (e.g position, time-stamp, etc.) with the already built nodes in the $d$ layers present in tree. After finding its way down the tree, the photo's information will be saved as a new leaf in the $D+1$ layer. Finding the position for each photo is the most important process of the DA. When all the photos are processed, one of the siblings (e.g. last entry) from each leaf in the $D+1$ layer as shown in Fig.~\ref{f9} will be forwarded to the event report requester.

It is worth to note that, in some cases, the optimal subset of photos cannot be obtained. In fact, the solution depends on the order of the processed photos. For example, if we have three pictures A, B, and C where A and B are detected similar in some way and B and C are also detected similar in another way. If the picture stream is A-B-C, then B will be discarded, but if the picture stream is B-A-C, then only B will be kept. It is difficult to obtain the optimal selection from the picture stream, known as NP-hard problem for a complete dataset.

\section{Experiments and Evaluation}\label{sec4}
In this section, we study the behavior of the photo type recognizer component as well as the data aggregator element toward different variations. We analyze each behavior and implement the models that provide the best performance.

\subsection{Photo Type Prediction}
\begin{figure}[t]
\begin{minipage}[h]{1\linewidth}
\vspace{0.4cm}
    \centering
    \includegraphics[width=8.75cm]{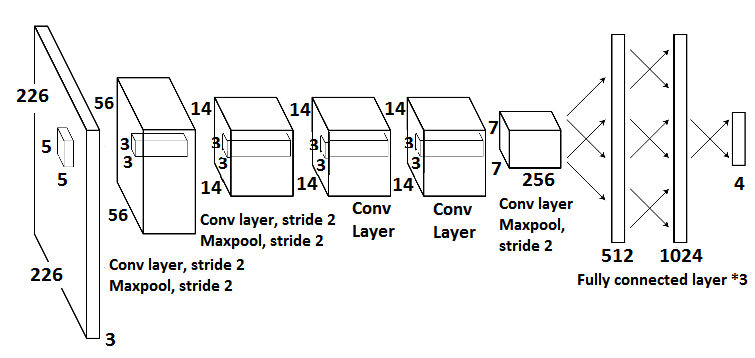}
    \caption{Customized 8-layer CNN.}
    \label{t1}
\end{minipage}\vspace{-0.3cm}
\end{figure}

We assume that we have $M=4$ classes, $3$ of them describe special events that interest the report requesters (fire, flood, and damaged infrastructure) and the fourth one represents a normal everyday event. For each event, we have collected $1000$ photos to build our dataset. Then, we study three different Convolutional Neural Network (CNN) models; 2 of them were based on ResNet-18~\cite{he2016deep} and the third one is a customized 8-layer CNN with 5 convolutional layers and 3 fully connected layers as shown in Fig. ~\ref{t1}. Moreover, the deep learning models which we approached were: 1) ResNet-18 fine-tuned, 2) ResNet-18 from scratch, and 3) a fully customized CNN. We used $K$-fold cross validation procedure with $K=10$ to split our dataset and improve the model performance.

\begin{figure}[!t]
\includegraphics[scale=0.321]{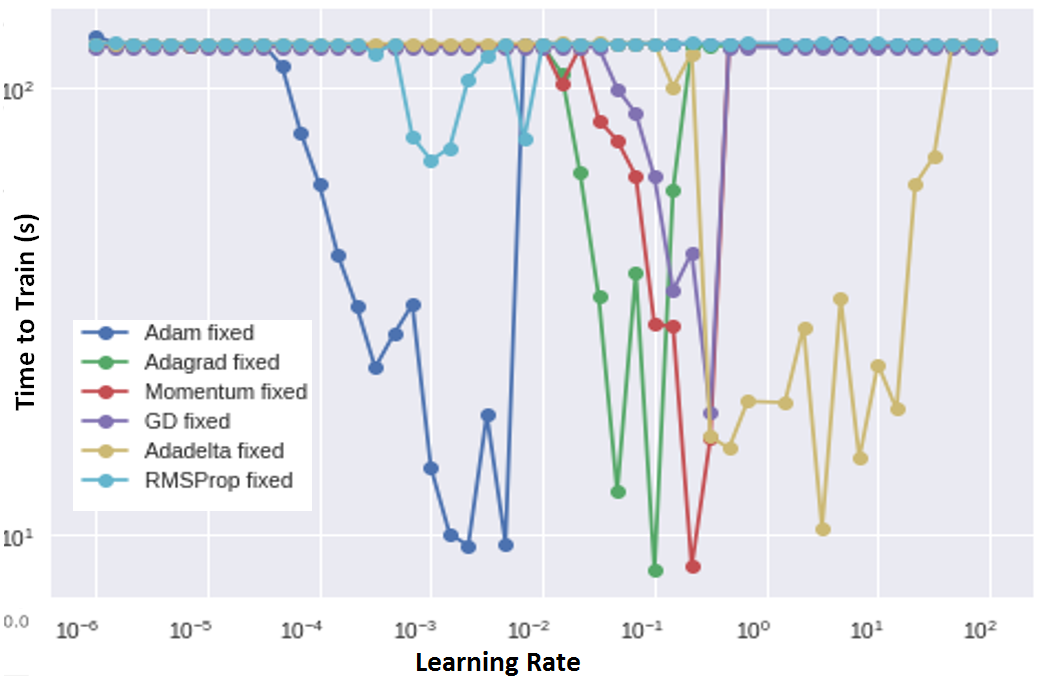}\vspace{-0.2cm}
 \caption{Time to Train vs Learning Rate, by Optimizer, for ResNet-18 from scratch.}
    \label{f2}\vspace{-0.4cm}
\end{figure}

\begin{figure}[t]
    \includegraphics[scale=0.335]{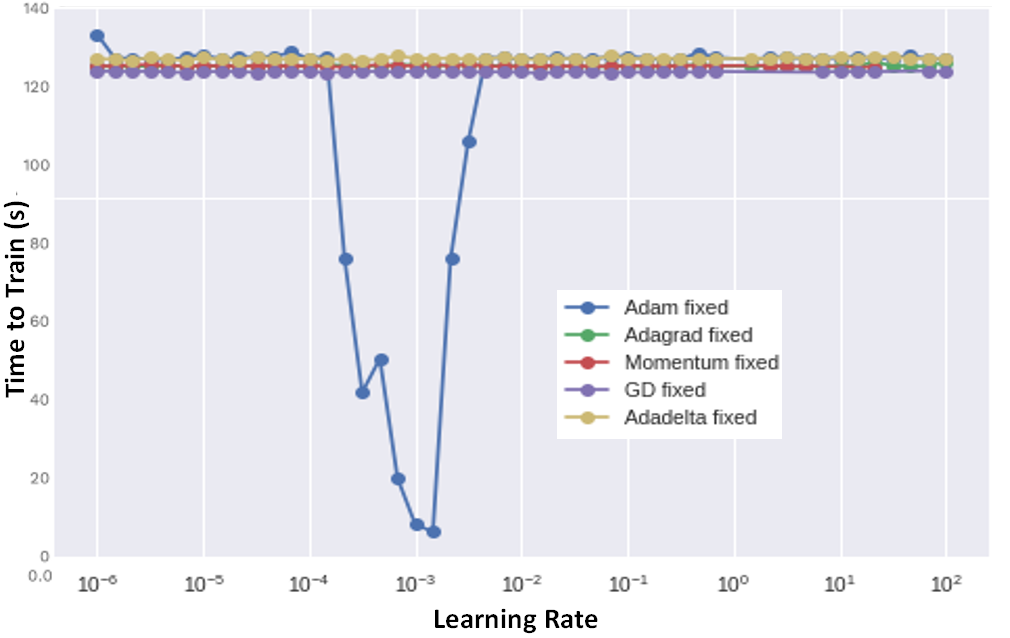}
   \caption{Time to Train vs Learning Rate, by Optimizer, for ResNet-18 fine-tuned model.}
    \label{f1}
\end{figure}

In order to speed up the training process and for the model to converge, we choose optimal independent variables by performing a training simulation for a batchset $=400$ equally composed of the 4 chosen classes. For the ResNet-18 fine-tuning, we consider that it should not take more than $120s$ to converge since it is already pretrained. However, for the ResNet-18 from scratch, and since the training starts with random weights, it will naturally takes more time to converge so we choose to set  $200s$ the threshold for model convergence. For both models, The Time to Train vs. learning rate, by the optimizer simulation as presented in Fig.~\ref{f2} and Fig.~\ref{f1} shows that the two models perform well in particular (learning rate, optimizer) combinations.

We implemented these two models along with the fully custom CNN model. The training of these three models were done using Google Colaboratory platform provided with the following specifications:\\
$\bullet$ GPU: 1x Tesla K80, compute 3.7, having 2496 CUDA cores, 12GB GDDR5 VRAM,\\
$\bullet$ CPU: 1x single core hyper threaded Xeon Processors @2.3Ghz,\\
$\bullet$ RAM: ~12.6 GB available.

\begin{table}[t]
\begin{center}
\caption{\label{p1} Deep Learning CNN photo recognition model's performance with Optimizer and Learning Rate (LR) as dependant variables}
\addtolength{\tabcolsep}{-4pt}
\resizebox{8.89cm}{!} {
\begin{tabular}{|c||c|c|c|c|c|c|c|c|}%
  \hline
 \textbf{} & \textbf{Optimizer} & \textbf{LR} & \textbf{AC (\%)
} & \textbf{PR (\%)} &  \textbf{RE (\%)}
 & \textbf{F1-S (\%)} & \textbf{LS} & \textbf{TtT (hours)} \\
  \hline
  \hline
\vtop{\hbox{\strut ResNet-18 }\hbox{\strut Fine-tuned}}
 & Adam Fixed  & 0.0012 & 91,32 &
\vtop{\hbox{\strut 0: 83 }\hbox{\strut 1: 91}
\hbox{\strut 2: 84}
\hbox{\strut 3: 79}}
&
\vtop{\hbox{\strut 0: 95}\hbox{\strut 1: 99}
\hbox{\strut 2: 94}
\hbox{\strut 3: 95}}

&
\vtop{\hbox{\strut 0: 88}\hbox{\strut 1: 94}
\hbox{\strut 2: 88}
\hbox{\strut 3: 86}}

&
0.4
&
72

\\
  \hline
\vtop{\hbox{\strut ResNet-18 }\hbox{\strut from scratch}} & Adagrad Fixed  & 0.1 & 95,14 & 
\vtop{\hbox{\strut 0: 92}\hbox{\strut 1: 97}
\hbox{\strut 2: 84}
\hbox{\strut 3: 98}}

&

\vtop{\hbox{\strut 0: 92}\hbox{\strut 1: 99}
\hbox{\strut 2: 99}
\hbox{\strut 3: 93}}
&

\vtop{\hbox{\strut 0: 92}\hbox{\strut 1: 97}
\hbox{\strut 2: 90}
\hbox{\strut 3: 95}}
&
1.21
&
95

 \\
  \hline
\vtop{\hbox{\strut Customized }\hbox{\strut model}}   & Adam Fixed  & 0.002 & 87,65 & 
\vtop{\hbox{\strut 0: 88}\hbox{\strut 1: 79}
\hbox{\strut 2: 74}
\hbox{\strut 3: 88}}

&
\vtop{\hbox{\strut 0: 77}\hbox{\strut 1: 95}
\hbox{\strut 2: 80}
\hbox{\strut 3: 93}}

&
\vtop{\hbox{\strut 0: 82}\hbox{\strut 1: 86}
\hbox{\strut 2: 76}
\hbox{\strut 3: 90}}

&
2.1
&
35

\\
 \hline
 \multicolumn{9}{l}{Accuracy (AC), Precision (PR), Recall (RE), F1-Score (F1-S), Loss (LS) and Time to train (TtT)} \\
 \multicolumn{9}{l}{as performance metrics. $[0..3]$ represent the one-hot encoding of the $4$ chosen classes.}\\
\end{tabular} }
\end{center}\vspace{-0.2cm}
\end{table}

\begin{figure}[!t]
\begin{minipage}[t]{1\linewidth} 
    \centering
    \includegraphics[width=1\textwidth]{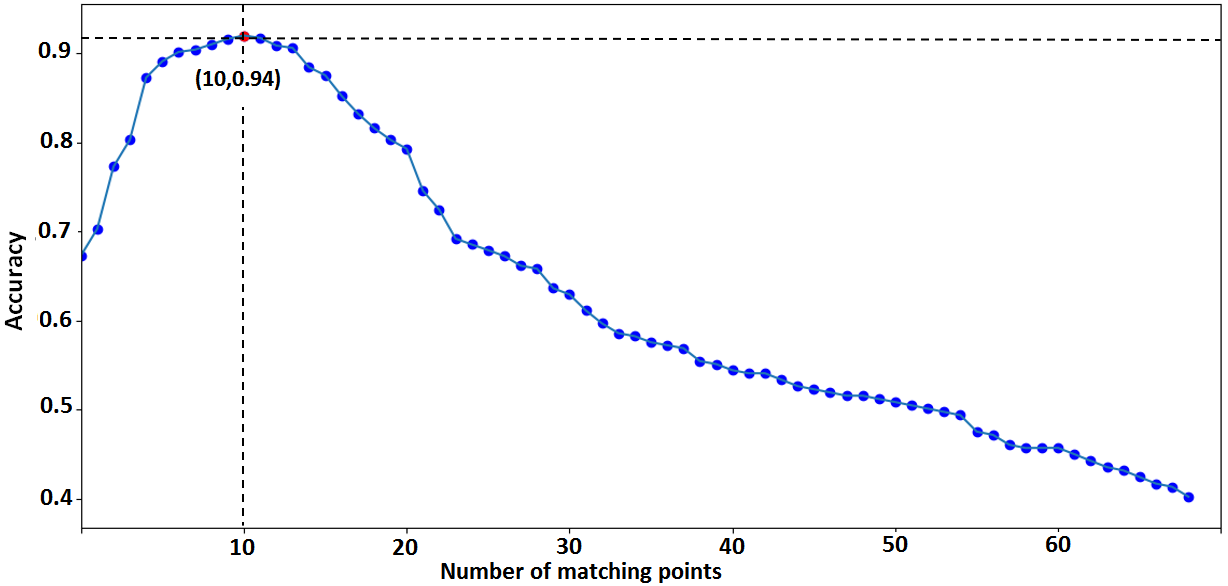}
    \caption{Accuracy of photo matching using SIFT while varing number of matching points.}
    \label{F55}
\end{minipage}\vspace{-0.5cm}    
\end{figure} 
As Table~\ref{p1} shows, both ResNet-18-based models perform well by giving an accuracy (AC) higher than 90\% and a loss (LS) lower than 1.5 unlike the customized model which achieves an accuracy close to 88\% with a loss higher than 2. However, the time to train (TtT) of the customized model is relatively lower compared to the two ResNet-18 models. This advantage can leverage the latter when it is the case of numerous classes. We notice that there is a trade-off between the two ResNet-18 models. On one hand, ResNet-18 from scratch has a high accuracy but also has a high loss. On the other hand, the fine-tuned version has a lower accuracy but its loss is lower.

\subsection{Data Aggregator}

We have implemented a 3-layer A-tree and tested the DA using 3 constraints: photo's time stamp, GPS coordinates, and features.
For the location matching, we used haversine formula in order to pair photo's geographical positions.\\
%\begin{equation}
 %d = 2  r \arcsin(\sqrt{\sin^{2}(\dfrac{\theta_{2}-\theta_{1})}{2}+ \cos \theta_{2} \cos \theta_{1}\sin^{2}(\dfrac{\phi_{2}-\phi_{1})}{2})} \tag{1}\label{eq:1}
%\end{equation}
To match photos’ visual and semantic features, we use the Scale-Invariant Feature Transform (SIFT). In order to determine an optimized number of keypoints for photo matching, we perform the simulation provided in Fig.~\ref{F55} where we measure the accuracy of the process while varying the number of reference matching points. We find out that matching 10 keypoints provide the highest accuracy.

\subsection{Framework Implementation}
\begin{figure}[t]
\begin{minipage}[t]{1\linewidth} 
\vspace{-0.2cm}
    \centering
    \includegraphics[scale=0.455]{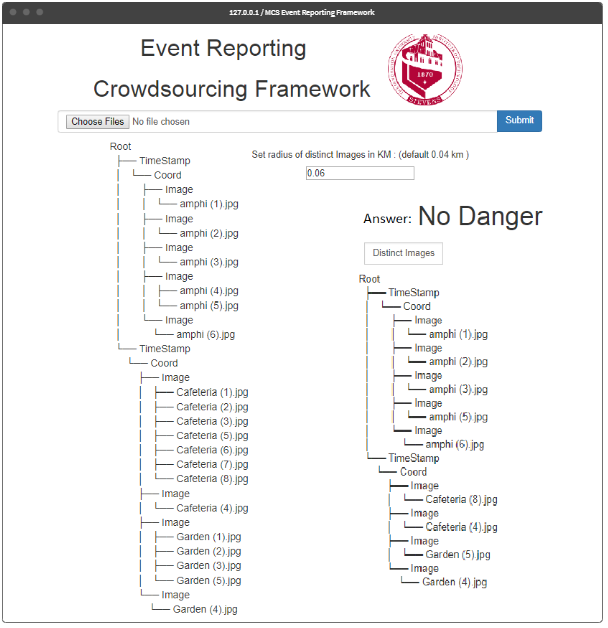}
   \caption{Event Reporting framework webview.}
    \label{f12}
\end{minipage}   \vspace{-0.4cm}  
\end{figure}

To prove the correctness, effectiveness, and robustness of the chosen  models,  tests  are  conducted  using  the MCS event reporting platform that we implemented on our local server. In the developed framework, we integrate the ResNet-18 fine-tuned model as well as a 4-layer A-tree shape hierarchical data structure. The server was developed using Python 3.7 and Flask 0.12.4 for the back-end. HTML, CSS, and JavaScript at the front-end. We test the framework using 19 photos taken within a university by a smartphone. Fig.~\ref{f12} shows the web view of the framework while highlighting the outputs of the PTP and DA phases.

\section{Conclusion and Future Work}\label{sec5}
In this paper, we developed a generic MCS framework for event reporting that supports photo filtering and near-optimal data selection for varied MCS tasks. Results of conducted evaluation have confirmed the effectiveness and efficiency
of our approach. Based on the findings in this paper, our future work will include the omitted phases of our framework mainly the task initiation in which we will study the dynamic arrival of workers. Also, we will consider using Edge Computing to reduce latency and data traffic. Finally , we  can extend our work by extending our framework to be data nature independent.
 \vspace{-0.4cm}  
\bibliographystyle{IEEEtran}
\bibliography{references}

\begin{thebibliography}{10}
\providecommand{\url}[1]{#1}
\csname url@samestyle\endcsname
\providecommand{\newblock}{\relax}
\providecommand{\bibinfo}[2]{#2}
\providecommand{\BIBentrySTDinterwordspacing}{\spaceskip=0pt\relax}
\providecommand{\BIBentryALTinterwordstretchfactor}{4}
\providecommand{\BIBentryALTinterwordspacing}{\spaceskip=\fontdimen2\font plus
\BIBentryALTinterwordstretchfactor\fontdimen3\font minus
  \fontdimen4\font\relax}
\providecommand{\BIBforeignlanguage}[2]{{%
\expandafter\ifx\csname l@#1\endcsname\relax
\typeout{** WARNING: IEEEtran.bst: No hyphenation pattern has been}%
\typeout{** loaded for the language `#1'. Using the pattern for}%
\typeout{** the default language instead.}%
\else
\language=\csname l@#1\endcsname
\fi
#2}}
\providecommand{\BIBdecl}{\relax}
\BIBdecl

\bibitem{8663367}
S.~{Dhingra}, R.~B. {Madda}, A.~H. {Gandomi}, R.~{Patan}, and M.~{Daneshmand},
  ``Internet of things mobile - air pollution monitoring system (iot-mobair),''
  \emph{to appear in IEEE Internet Things J.}, 2019.

\bibitem{8053134}
M.~{Habibzadeh}, W.~{Xiong}, M.~{Zheleva}, E.~K. {Stern}, B.~H. {Nussbaum}, and
  T.~{Soyata}, ``Smart city sensing and communication sub-infrastructure,'' in
  \emph{IEEE Int. Midwest Symp. Circuits Syst (MWSCAS'17)}, Boston, MA, USA,
  Aug. 2017.

\bibitem{7807311}
H.~{Chen}, B.~{Guo}, Z.~{Yu}, L.~{Chen}, and X.~{Ma}, ``A generic framework for
  constraint-driven data selection in mobile crowd photographing,'' \emph{IEEE
  Internet of Things J.}, vol.~4, no.~1, pp. 284--296, Feb. 2017.

\bibitem{6871668}
D.~{Zhang}, L.~{Wang}, H.~{Xiong}, and B.~{Guo}, ``4w1h in mobile crowd
  sensing,'' \emph{IEEE Commun. Mag.}, vol.~52, no.~8, pp. 42--48, Aug. 2014.

\bibitem{Kim:2011:CWP:1978942.1979251}
S.~Kim, C.~Robson, T.~Zimmerman, J.~Pierce, and E.~M. Haber, ``Creek watch:
  Pairing usefulness and usability for successful citizen science,'' in
  \emph{SIGCHI Conf. Human Factors Comput Systs. (CHI'11)}, Vancouver, BC,
  Canada, May 2011.

\bibitem{Koukoumidis:2011:SLM:1999995.2000008}
E.~Koukoumidis, L.-S. Peh, and M.~R. Martonosi, ``Signalguru: Leveraging mobile
  phones for collaborative traffic signal schedule advisory,'' in \emph{Int.
  Conf. Mobile. Syst., App., Services. (MobiSys '11)}, New York, NY, USA, June
  2011.

\bibitem{Guo:2014:FCP:2638728.2638730}
B.~Guo, H.~Chen, Z.~Yu, X.~Xie, S.~Huangfu, and Z.~Wang, ``Fliermeet:
  Cross-space public information reposting with mobile crowd sensing,'' in
  \emph{IEEE Trans. Mobile Comput.}, vol.~14, pp. 2020--2033, Oct. 2015.

\bibitem{mobishop}
S.~Sehgal and C.~T.~C. Salil S.~Kanhere, ``Mobishop: Using mobile phones for
  sharing consumer pricing information,'' in \emph{Demo Session Intl. Conf.
  Distrib. Comput. Sensor Syst (DCOSS'08)}, New York, NY, USA, June 2008.

\bibitem{7524481}
H.~{Chen}, B.~{Guo}, Z.~{Yu}, and Q.~{Han}, ``Toward real-time and cooperative
  mobile visual sensing and sharing,'' in \emph{IEEE Int. Conf. Compt. Commun.
  (INFOCOM'16)}, San Francisco, CA, USA, April, 2016.

\bibitem{Tuite:2011:PTE:1978942.1979146}
K.~Tuite, N.~Snavely, D.-y. Hsiao, N.~Tabing, and Z.~Popovic, ``Photocity:
  Training experts at large-scale image acquisition through a competitive
  game,'' in \emph{SIGCHI Conf. Human Factors Comput Systs. (CHI’11)},
  Vancouver, BC, Canada, May 2011.

\bibitem{White:2011:WAT:1997768.1997797}
J.~White, C.~Thompson, H.~Turner, B.~Dougherty, and D.~C. Schmidt,
  ``Wreckwatch: Automatic traffic accident detection and notification with
  smartphones,'' \emph{Mob. Netw. Appl (MONET'11).}, vol.~16, pp. 285--303,
  Secaucus, NJ, USA, June 2011.

\bibitem{he2016deep}
K.~He, X.~Zhang, S.~Ren, and J.~Sun, ``Deep residual learning for image
  recognition,'' in \emph{IEEE Conf. Comp. Vis. Pattern. Recog (CVPR'18).},
  Utah, USA, June, 2016.

\end{thebibliography}
\end{document}